\documentclass[letterpaper, 10 pt, conference]{ieeeconf}  % Comment this line out if you need a4paper

\IEEEoverridecommandlockouts                              % This command is only needed if 
\usepackage{amsmath,amssymb,amsfonts,amsthm}

\usepackage{algorithmic}
\usepackage{graphicx}
\usepackage{algorithm,algorithmic}
\usepackage{hyperref}
\hypersetup{hidelinks}
\usepackage{textcomp}
\usepackage{mathtools}
\usepackage{pgfplots}
\usepackage{cancel}
\usepackage[dvipsnames]{xcolor}

\usepackage[backend=biber,style=ieee]{biblatex}
\let\labelindent\relax
\usepackage{enumitem}

\pgfplotsset{compat=1.18}

\newcommand{\bbE}{\mathbb{E}}

\newcommand{\calN}{\mathcal{N}}

\newcommand{\ggl}{g_\textnormal{l}}

\newcommand{\bargge}{\bar{g}_\textnormal{e}}
\newcommand{\barggi}{\bar{g}_\textnormal{i}}
\newcommand{\dotxxe}{{\dot x}_\textnormal{e}}
\newcommand{\dotxxi}{{\dot x}_\textnormal{i}}
\newcommand{\xxe}{{x}_\textnormal{e}}
\newcommand{\xxi}{{x}_\textnormal{i}}
\newcommand{\EEe}{{E}_\textnormal{e}}
\newcommand{\EEi}{{E}_\textnormal{i}}
\newcommand{\psie}{{\psi}_\textnormal{e}}
\newcommand{\psii}{{\psi}_\textnormal{i}}
\newcommand{\epse}{{\varepsilon}_\textnormal{e}}
\newcommand{\epsi}{{\varepsilon}_\textnormal{i}}
\newcommand{\aae}{{a}_\textnormal{e}}
\newcommand{\aai}{{a}_\textnormal{i}}
\newcommand{\bbe}{{b}_\textnormal{e}}
\newcommand{\bbi}{{b}_\textnormal{i}}
\newcommand{\sigmae}{{\sigma}_\textnormal{e}}
\newcommand{\sigmai}{{\sigma}_\textnormal{i}}
\newcommand{\xie}{{\xi}_\textnormal{e}}
\newcommand{\xii}{{\xi}_\textnormal{i}}
\newcommand{\Ifast}{{I}_\textnormal{fast}}
\newcommand{\Islow}{{I}_\textnormal{slow}}
\newcommand{\Is}{{I}_\textnormal{s}}
\newcommand{\vrest}{{v}_\textnormal{rest}}
\newcommand{\vABE}{{v}_\textnormal{ABE}}
\newcommand{\vth}{{v}_\textnormal{th}}

\newtheorem{proposition}{Proposition}
\newtheorem{definition}{Definition}

\title{\LARGE \bf
Neuromodulation Through Intrinsic Noise
}

\author{Sunghyun Koo$^{1}$, Jin Gyu Lee$^{1}$, and Rodolphe Sepulchre$^{2}$% <-this % stops a space
\thanks{*The research leading to these results has received funding from the European Research Council under the Advanced ERC Grant Agreement SpikyControl n.101054323.}% <-this % stops a space
\thanks{$^{1}$Sunghyun Koo and Jin Gyu Lee are with ASRI, Department of Electrical and Computer Engineering, Seoul National University, Seoul 08826, South Korea
        {\tt\small shkoo@cdsl.kr}, {\tt\small jingyu.lee@snu.ac.kr}}%
\thanks{$^{2}$Rodolphe Sepulchre is with STADIUS, Department of Electrical Engineering (ESAT), KU Leuven, 3001 Leuven, Belgium, and Department of Engineering, University of Cambridge, Cambridge CB2 1PZ, United Kingdom
        {\tt\small rodolphe.sepulchre@kuleuven.be}}%
}

\begin{document}

\maketitle
\thispagestyle{empty}
\pagestyle{empty}

%%%%%%%%%%%%%%%%%%%%%%%%%%%%%%%%%%%%%%%%%%%%%%%%%%%%%%%%%%%%%%%%%%%%%%%%%%%%%%%%
\begin{abstract}
Neuromodulation, the modulation of excitability through the regulation of neuronal parameters, is a key tool in neuromorphic control design.
This paper investigates how spiking excitability can be modulated by intrinsic noise.
Focusing on a simplified memristive neuron model, we show that intrinsic noise plays distinct functional roles depending on the feedback pathways:
Intrinsic noise in the fast excitatory feedback shapes the nature of excitability and modulates the sensitivity of the neuron to external/internal stimuli, whereas that in the slow inhibitory feedback provides internal stimuli that can trigger stochastic spiking.
We expect these findings to guide the design of neuromorphic systems that harness intrinsic noise to exploit beneficial stochastic phenomena, such as stochastic resonance.
\end{abstract}
%%%%%%%%%%%%%%%%%%%%%%%%%%%%%%%%%%%%%%%%%%%%%%%%%%%%%%%%%%%%%%%%%%%%%%%%%%%%%%%%
\section{Introduction}
\label{sec:introduction}

Animal nervous systems are, arguably, among the most sophisticated control systems on Earth.
Their event-driven, spike-based operation supports efficient and robust interactions with dynamic environments \cite{ribar21csm,sepulchre26csm}.
These capabilities have motivated the development of neuromorphic control systems, which seek to incorporate principles of neural computation into engineered controllers \cite{jang25arxiv,petri25cdc,huo26}.
A key step toward this goal is to understand how neuronal spiking behaviors can be shaped and modulated \cite{ribar21csm,burger25cdc,wroblowski2026ecc}.

\begin{figure}
    \centering
    \includegraphics[width=\linewidth]{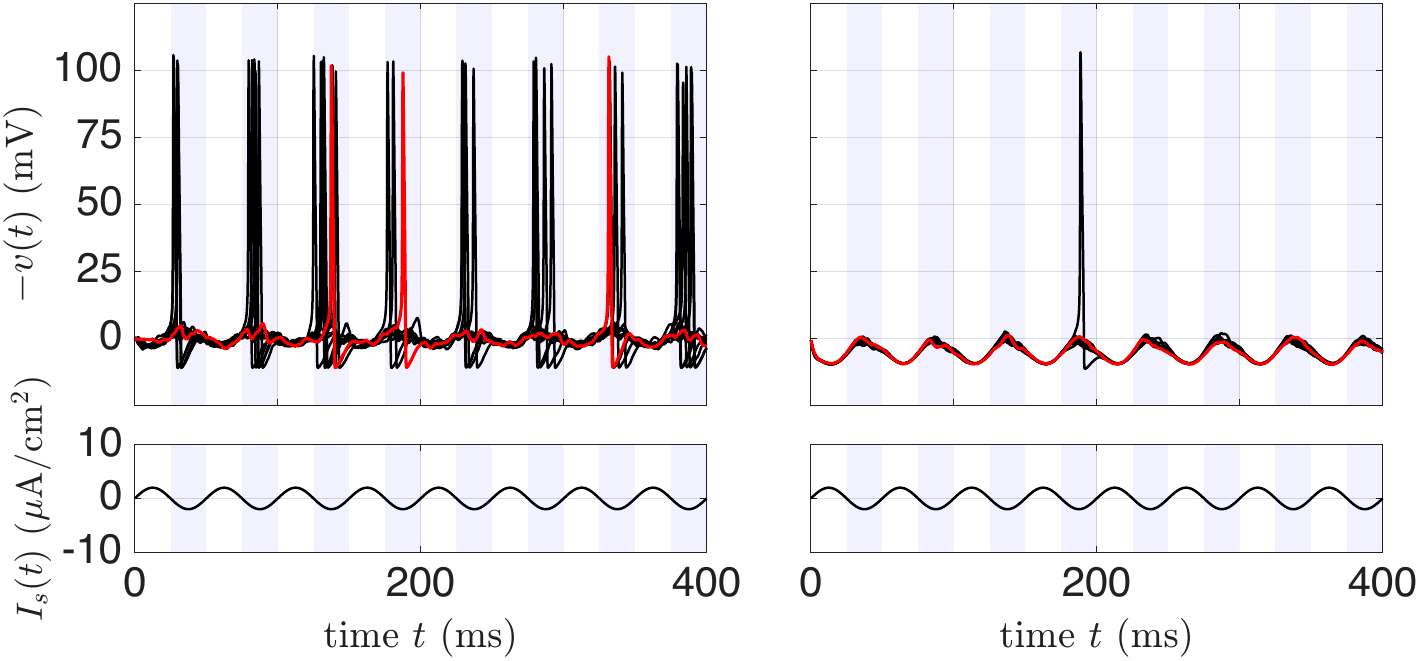}
    \caption{Responses of the Hodgkin-Huxley model with intrinsic noise to the subthreshold input current $\Is(t) = 2 \sin(2\pi t/50)\,\mathrm{\mu A/cm^2}$. 
    Each panel shows 10 repeated trials for a different set of parameters, with one trajectory highlighted in red.
    The shaded regions indicate intervals during which the input current is negative. Left: Parameters are taken from \cite[Table.~3]{HH}. Right: The leak equilibrium potential is replaced with $V_l=2.72 \, \mathrm{mV}$.}
    \label{fig:HH}
\end{figure}

While most efforts in this direction have focused on deterministic settings, biological neurons are inherently stochastic due to the random opening and closing of ion channels.
Importantly, such noise is not merely a disturbance; \emph{neural systems can harness this stochasticity as a computational resource} \cite{faisal2008nature,mainen1995science}.
A relevant example is \emph{stochastic resonance}, in which noise enhances the detection of a subthreshold signal under suitable conditions \cite{gammaitoni1998}.
Fig.~\ref{fig:HH} illustrates this effect using Hodgkin-Huxley neurons with two parameter sets; with the parameters yielding resonance, the neurons fire in synchrony with the subthreshold input, whereas with the other parameters, their membrane potentials merely fluctuate.

To inform the design of neuromorphic systems that leverage these benefits, this paper investigates how noise modulates neuronal dynamics.
We consider a memristive neuron model consisting of two feedback pathways \cite{burger25cdc} and incorporate intrinsic noise into each pathway.
Based on this model, we first examine the effects of noise on excitability.
Qualitatively, we show that noise can either eliminate excitability or enable anode break excitation.
Quantitatively, we discuss how noise affects sensitivity, i.e., the propensity of a neuron to spike.
We also investigate stochastic spiking, an irregular firing behavior induced by noise.
Qualitatively, we relate its occurrence to anode break excitability.
Quantitatively, we analyze how its rate depends on the noise intensities and how it affects the detection of subthreshold signals.
In particular, we show that intrinsic noise in the two feedback pathways contributes to stochastic resonance through distinct yet complementary mechanisms.

The rest of the paper is organized as follows.
Section~\ref{sec:memristive_model} introduces the neuron model considered in this paper.
Section~\ref{sec:modulation_of_excitability} investigates how intrinsic noise modulates excitability.
Section~\ref{sec:modulation_of_stochastic_spiking} examines how intrinsic noise modulates stochastic spiking, including an analysis of stochastic resonance.
Section~\ref{sec:concluding_remarks} concludes the paper.

\section{Memristive Model of Neuron}
\label{sec:memristive_model}

In this section, we first review the memristive model proposed in \cite{burger25cdc}.
We then present a graphical framework for interpreting its excitability and introduce a stochastic extension of the model that incorporates intrinsic noise.

\subsection{Review of the Memristive Model}

In \cite{burger25cdc}, the following model of a neuron is proposed:
\begin{subequations}
\label{eqn:memristive_model}
    \begin{align}
        C\dot v &= \!- \ggl v + \bargge \psie(\xxe)(\EEe\!-\!v)\! + \barggi \psii (\xxi) (\EEi\!-\!v)\! + \Is \label{eqn:v_dynamics}\\
        \dotxxe &= {\epse}^{-1} (v-\xxe) \label{eqn:e_memristor}\\
        \dotxxi &= {\epsi}^{-1} (v-\xxi) \label{eqn:i_memristor}
    \end{align}
\end{subequations}
While its conductance-based formulation offers a more direct biophysical interpretation than phenomenological models such as Izhikevich's \cite{izhikevich}, its simpler structure facilitates analysis relative to the Hodgkin-Huxley model \cite{HH}.

The dynamics of the membrane potential $v$ is modeled as \eqref{eqn:v_dynamics}.
Here, $C$ denotes the capacitance of the neuron, and the right-hand side consists of four current sources, namely, a leak current with conductance $\ggl$, an excitatory current (\emph{e-current}) with conductance $\bargge \psie(\xxe)$ and equilibrium potential $\EEe>0$, an inhibitory current (\emph{i-current}) with conductance $\barggi \psii(\xxi)$ and equilibrium potential $\EEi<0$, and an external (synaptic) current~$\Is$.
The conductances for e- and i-currents are memductances with respective activation functions $\psie$ and $\psii$ and states $\xxe$ and $\xxi$.
We adopt sigmoidal activation functions
\begin{equation*}
    \psie(\xxe) = \Phi\!\left(\frac{\xxe-\aae}{\bbe}\right)\!,\quad \psii(\xxi) = \Phi\!\left(\frac{\xxi-\aai}{\bbi}\right)\!,
\end{equation*}
where $\Phi$ denotes the standard Gaussian cumulative distribution function (CDF), $\aae,\aai\in[\EEi,\EEe]$ specify the locations of activation, and $\bbe,\bbi>0$ determine their sensitivity.
The dynamics of memductance states are given by \eqref{eqn:e_memristor} and \eqref{eqn:i_memristor}, where $\epse$ and $\epsi$ denote their respective time constants; we assume $\epse \ll 1 \ll \epsi$ as in biological neurons.

The current sources, together with the timescale separation, lead to excitability under suitable parameters.
The fast e-current renders the dynamics of the system bistable on the fast timescale.
A brief excitatory pulse thus makes the membrane potential jump to the excited potential, generating a spike.
We shall call this bistability \emph{memory}.
On the slow timescale, the slow current is gradually activated, and at a certain point, the memory disappears.
After that, the membrane potential will recover to the resting potential.

\subsection{Graphical Interpretation of the Excitability}
\label{sec:graphical_interpretation}

\begin{figure}
    \centering
    \includegraphics[width=0.93\linewidth]{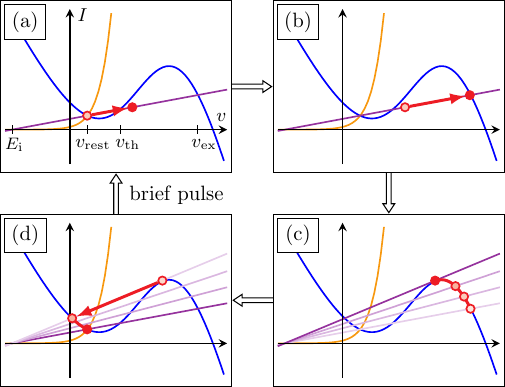}
    \caption{Graphical interpretation of the excitability. The blue, yellow, and purple curves represent $\Ifast(v)$, $\Islow(v;v)$, and $\Islow(v;\xxi)$, respectively.}
    \label{fig:excitability}
\end{figure}

We here propose a graphical interpretation of the mechanism of excitability.
Since $\epse \ll 1$, the state of the fast memductance immediately follows the membrane potential.
Thus, in view of Tikhonov's theorem, \eqref{eqn:memristive_model} can be reduced to
\begin{equation}
\label{eqn:reduced_model}
\begin{aligned}
    C\dot v &= \Ifast(v) - \Islow(v;\xxi) + \Is,\\
    \dotxxi &= \epsi^{-1} (v-\xxi),
\end{aligned}
\end{equation}
where we denote
\begin{align*}
    &\Ifast(v) \coloneqq -\ggl v + \bargge \psie(v)(\EEe-v),\\
    &\Islow(v;\xxi) \coloneqq -\barggi \psii(\xxi) (\EEi-v).
\end{align*}
We will work with three curves, $\Ifast(v)$, $\Islow(v;v)$, and $\Islow(v;\xxi)$.
As shown in Fig.~\ref{fig:excitability}, $\Ifast(v)$ has a reverse-N profile that gives rise to the memory.
Following Ohm's law, $\Islow(v;\xxi)$ varies linearly with $v$ for each $\xxi$.

\begin{itemize}
    \item In the absence of external input ($\Is=0$), the membrane potential remains at the resting potential $\vrest$, which is determined by $\Ifast(\vrest)=\Islow(\vrest;\vrest)$, i.e., $\vrest$ is the intersection of $\Ifast(v)$ and $\Islow(v;v)$.
    \item 
    Suppose that $\Ifast(v)$ and $\Islow(v;\vrest)$ intersect at three points, with the resting potential $\vrest$ as the leftmost one (Fig.~\ref{fig:excitability}(a)--(b)). The other two intersections correspond to the threshold $\vth$ and excited potential $v_\text{ex}$. These are the equilibrium points on the fast timescale.
    \item Once a brief excitatory pulse brings the membrane potential above the threshold (Fig.~\ref{fig:excitability}(a)), the fast dynamics carries it toward the excited potential (Fig.~\ref{fig:excitability}(b)) because $\Ifast(v) > \Islow(v;\xxi)$ over this voltage range.
    \item On the slow timescale, $\xxi$ gradually increases toward $v$, causing the slope of $\Islow(v;\xxi)$ to increase. The membrane potential tracks the intersection of $\Ifast(v)$ and $\Islow(v;\xxi)$ (Fig.~\ref{fig:excitability}(c)).
    \item Once $\Ifast(v)$ and $\Islow(v;\xxi)$ become tangent, the fast dynamics makes the membrane potential hyperpolarized since $\Ifast(v)< \Islow(v;\xxi)$. Then, $\xxi$ decreases toward $v$ on the slow timescale, and the membrane potential gradually recovers its resting potential (Fig.~\ref{fig:excitability}(d)).
\end{itemize}

Motivated by this graphical interpretation, we introduce a graphical definition of excitability.

\begin{definition}
\label{def:excitable}
    The memristive neuron \eqref{eqn:memristive_model} or its reduced model \eqref{eqn:reduced_model} is said to be \emph{excitable} if
    \begin{itemize}[nosep]
        \item $\Ifast(v)$ and $\Islow(v;v)$ intersect at exactly one point, which we refer to as the resting potential $\vrest$,
        \item $\Ifast(v)$ and $\Islow(v;\vrest)$ intersect at exactly three points, with $\vrest$ as the leftmost intersection.
    \end{itemize}
\end{definition}

\subsection{Memristive Model with Intrinsic Noise}

In biological neurons, the stochastic opening and closing of ion channels give rise to fluctuations in ionic conductances.
To model these, we add white noise to the dynamics of the memductance states as follows:
\begin{subequations}
\label{eqn:stochastic_memristive_model}
\begin{align}
    C\dot v &=\! -\ggl v + \bargge \psie(\xxe)(\EEe\!-\!v)\! + \barggi \psii (\xxi) (\EEi\!-\!v)\! + \Is \label{eqn:stochastic_v_dynamics}\\
    \dotxxe &= {\epse}^{-1} (v-\xxe) + {\epse^{-1/2}\sigmae} \xie \label{eqn:stochastic_e_memristor}\\
    \dotxxi &= {\epsi}^{-1} (v-\xxi) + \epsi^{-1/2}{\sigmai} \xii \label{eqn:stochastic_i_memristor}
\end{align}
\end{subequations}
Here, $\sigmae$ and $\sigmai$ denote noise intensities for each memductance, and $\xie$ and $\xii$ are independent white noise processes.
The noise intensities scale with the inverse square roots of the time constants, following the derivation in \cite{yu2017physicaA}.
Because these noise sources originate within the neuron, such noise is called \emph{intrinsic noise}. Specifically, we refer to the noise associated with the e-current and i-current as \emph{e-noise} and \emph{i-noise}, respectively.

Timescale separation still allows us to derive a reduced model, but doing so requires stochastic averaging \cite{wainrib2012,franovic2020chaos,genadot2012}.
For a fixed membrane potential $v$, the process $\xxe$ governed by \eqref{eqn:stochastic_e_memristor} has the stationary distribution $\calN(v,\sigmae^2/2)$ regardless of $\epse$.
Because $\xxe$ evolves on a faster timescale than $v$ and $\xxi$, averaging over this stationary distribution gives the reduced model \cite[Sec.~7.9]{Freidlin}
\begin{equation}
\label{eqn:noisy_reduced_model}
\begin{aligned}
    C\dot v &= \Ifast(v;\sigmae) - \Islow(v;\xxi) + \Is,\\
    \dotxxi &= \epsi^{-1} (v-\xxi) + \epsi^{-1/2}\sigmai \xii,
\end{aligned}
\end{equation}
where $\Islow(v;\xxi)$ is defined as in the deterministic case and
\begin{align*}
    \Ifast(v;\sigmae) \coloneqq -\ggl v + \bargge \underbrace{\bbE_{\xxe\sim\calN(v,\sigmae^2/2)}[ \psie(\xxe)]}_{\eqqcolon\psie(v;\sigmae)}(\EEe-v),
\end{align*}
which coincides with $\Ifast(v)$ when $\sigmae = 0$.
Denoting by $\phi_{\sigmae^2/2}$ the probability distribution function (PDF) of $\calN(0,\sigmae^2/2)$, one can observe that
\begin{equation}
\label{eqn:convolution}
    \psie(v;\sigmae) = (\psie \ast \phi_{\sigmae^2/2})(v) = \Phi\!\left(\frac{v-\aae}{\sqrt{\bbe^2+\sigmae^2/2}}\right),
\end{equation}
where $\ast$ denotes the convolution operator.

Based on \eqref{eqn:convolution}, the effect of e-noise can be understood as smoothing the activation function of the e-current.
Specifically, a stronger e-noise produces a smoother averaged activation function $\psi_e(\,\cdot\,;\sigmae)$ because the corresponding density function $\phi_{\sigmae^2/2}$ is more widely spread.

\section{Modulation of Excitability}
\label{sec:modulation_of_excitability}

In this section, we investigate how e-noise qualitatively and quantitatively modulates the excitability.
We here assume $\sigmai=0$ to isolate the effect of e-noise.
The reduced model \eqref{eqn:noisy_reduced_model} is then deterministic.
(We consider i-noise in Section~\ref{sec:modulation_of_stochastic_spiking}.)

\subsection{Elimination of Excitability}

\begin{figure}
    \centering
    \includegraphics[width=0.6\linewidth]{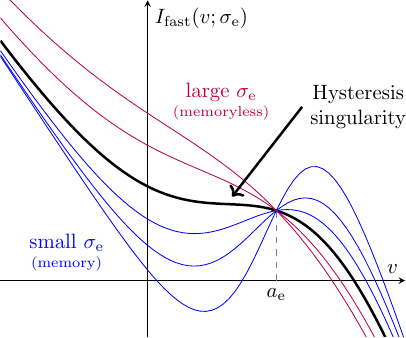}
    \caption{The current component $\Ifast(v;\sigmae)$ undergoes a hysteresis singularity as the noise intensity $\sigmae$ varies.}
    \label{fig:memory}
\end{figure}

Recall that the source of excitability is the memory on the fast timescale,
which is characterized by the reverse-N-shaped profile of $\Ifast$.
In view of \eqref{eqn:convolution}, e-noise smooths out this nonlinearity, and increasing its intensity may eventually lead to a loss of memory and excitability, as shown in Fig.~\ref{fig:memory}.
Indeed, we prove the following proposition.

\begin{proposition}
\label{prop:memory}
    Suppose that $\Ifast(v;0)$ is reverse-N-shaped, i.e., ${\partial_v \Ifast(v;0)} = 0$ has exactly two distinct solutions.
    Suppose that $\ggl/\bargge > -\Phi(-\sqrt{2}) +\sqrt{2} \Phi'(-\sqrt{2})\approx 0.1289$.
    Then, there exists $\sigmae^\ast >0$ such that ${\partial_v \Ifast(v;\sigmae)} = 0$ has
    \begin{itemize}[nosep]
        \item exactly two distinct solutions if $0\leq\sigmae < \sigmae^\ast$,
        \item a unique solution if $\sigmae = \sigmae^\ast$,
        \item no solution if $\sigmae > \sigmae^\ast$.
    \end{itemize}
    Moreover, the case $\sigmae=\sigmae^\ast$ corresponds to the hysteresis singularity in that the unique solution $v^\ast$ of ${\partial_v \Ifast(v;\sigmae^\ast)} = 0$ satisfies ${\partial_v^2 \Ifast(v^\ast;\sigmae^\ast)} = 0$ and ${\partial_v^3 \Ifast(v^\ast;\sigmae^\ast)} < 0$.
\end{proposition}

\begin{proof}
    Let $\phi$ denote the standard Gaussian PDF, and denote
    \begin{equation*}
        w \coloneqq \frac{v-\aae}{\sqrt{\bbe^2+\sigmae^2/2}}, \quad r \coloneqq \frac{\EEe-\aae}{\sqrt{\bbe^2+\sigmae^2/2}}.
    \end{equation*}
    Then, one can verify that
    \begin{equation*}
        \frac{\partial \Ifast (v;\sigmae)}{\partial v} = -\ggl - \bargge(\Phi(w) +  \phi(w) (w-r))
    \end{equation*}
    The number of solutions of $\partial_v \Ifast(v;\sigmae)=0$ thus coincides with that of
    \begin{equation}
    \label{eqn:h_equation}
        h(w;r) \coloneqq \Phi(w)  + \phi(w)(w-r) = - {\ggl}/{\bargge}.
    \end{equation}

    Noting that $\phi'(w) = -w\phi(w)$, one can observe that
    \begin{equation*}
        \frac{\partial h(w;r)}{\partial w} = (2 -w(w-r))\phi(w).
    \end{equation*}
    It can also be seen that $h(w;r) \to 0$ as $w\to -\infty$ and $h(w;r)\to 1$ as $w\to\infty$.
    Combining these, we can conclude that $h(w;r)$ attains its minimum and maximum at $w_1(r)\coloneqq \frac{r-\sqrt{r^2+8}}{2}$ and $w_2(r)\coloneqq \frac{r+\sqrt{r^2+8}}{2}$, respectively, and \eqref{eqn:h_equation} has
    \begin{itemize}[nosep]
        \item exactly two solutions if $h(w_1(r);r)<-\ggl/\bargge$,
        \item a unique solution if $h(w_1(r);r)=-\ggl/\bargge$,
        \item no solution if $h(w_1(r);r) > -\ggl/\bargge$.
    \end{itemize}
    In particular, $h(w_1(r_0);r_0)<-\ggl/\bargge$ for $r_0 \coloneqq (\EEe-\aae)/\bbe$ because $\partial_v\Ifast(v;0)=0$ has two distinct solutions.

    Since $r$ decreases as $\sigmae$ increases (note that $\EEe>\aae$) and
    \begin{align*}
        \frac{d h(w_1(r);r)}{d r}
        &= \underbrace{\frac{\partial h(w_1(r);r)}{\partial w}}_{=0} \frac{dw_1(r)}{dr} +\underbrace{\frac{\partial h(w_1(r);r)}{\partial r}}_{=-\phi(w_1(r))} < 0,
    \end{align*}
    increasing $\sigmae$ from 0 to $\infty$ causes $h(w_1(r);r)$ to increase monotonically from $h(w_1(r_0);r_0)$ to $h(w_1(0);0)=\Phi(-\sqrt{2})-\sqrt{2}\phi(\sqrt{2})$.
    Since $h(w_1(r_0);r_0)<-\ggl/\bargge<h(w_1(0);0)$, the first claim is proved.

    The second claim can be verified by direct computation with $\sigmae=\sigmae^\ast$ and $v^\ast = w_1(r)$.
\end{proof}

Proposition~\ref{prop:memory} implies that $\Ifast(v;\sigmae)$ retains its reverse-N-shaped profile for $\sigmae<\sigmae^\ast$, whereas it becomes monotonically decreasing for $\sigmae>\sigmae^\ast$.
Consequently, when $\sigmae>\sigmae^\ast$, the memristive neuron~\eqref{eqn:stochastic_memristive_model}, or its reduced model \eqref{eqn:noisy_reduced_model}, cannot exhibit excitability for any choice of i-current parameters.

\subsection{Emergence of Anode Break Excitability}
\label{sec:rebound}

While e-noise degrades memory and eventually eliminates excitability, an intermediate intensity of e-noise can give rise to anode break excitability (ABE).
This section analyzes how this noise-induced ABE emerges.
To make this notion precise, we adopt the following definition.

\begin{definition}
    An excitable memristive neuron governed by \eqref{eqn:noisy_reduced_model} is said to be \emph{anode break excitable} if there exists $\vABE\leq\vrest$ such that, for each $v_0<\vABE$, the trajectory initialized at $v(0)=\xxi(0)=v_0$ produces a spike under the assumption that $\sigmai=0$.\footnote{When $\sigmai>0$, anode break excitation can randomly be skipped.}
    In other words, the neuron fires a spike upon removal of an inhibitory external current that has driven it to a steady state with the membrane potential below a certain threshold $\vABE$.
\end{definition}

We next characterize ABE using the graphical framework developed in Section~\ref{sec:graphical_interpretation}.

\begin{proposition}
\label{prop:characterization_of_rebound_excitability}
    Consider a memristive neuron governed by \eqref{eqn:noisy_reduced_model}.
    Suppose that it is excitable.
    \begin{itemize}[nosep]
        \item It exhibits ABE under sufficient timescale separation if the graph of $\Ifast(v;\sigmae)$ intersects the $v$-axis only once.
        \item It does not exhibit ABE under sufficient timescale separation if the graph of $\Ifast(v;\sigmae)$ intersects the $v$-axis at three points.
    \end{itemize}
\end{proposition}

\begin{figure}
    \centering
    \includegraphics[width=0.93\linewidth]{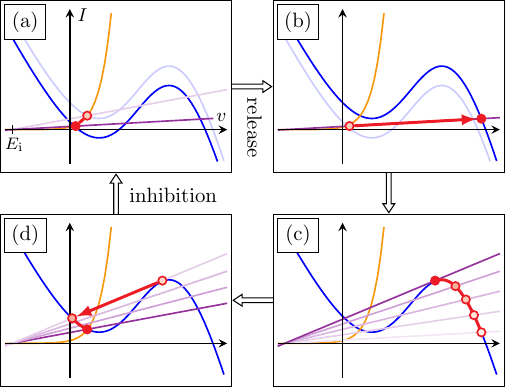}
    \caption{Mechanism of anode break excitation. The blue curve represents $\Ifast(v;\sigmae)+\Is$, and the other curves are defined as in Fig.~\ref{fig:excitability}.}
    \label{fig:rebound}
\end{figure}

\begin{proof}
    We first clarify the mechanism of anode break excitation using the proposed graphical framework.
    \begin{itemize}[nosep]
        \item Suppose that an inhibitory external current $\Is<0$ is applied, and the neuron has reached a steady state with $v=\xxi=v_0<\vrest$ (Fig.~\ref{fig:rebound}(a)). Suppose further that $\Ifast(v;\sigmae)$ and $\Islow(v;v_0)$ intersect at exactly one point.

        \item 
        When the inhibitory external current is removed, the fast dynamics carries the membrane potential to the excited potential because $\Ifast(v;\sigmae) > \Islow(v;v_0)$ in this voltage range (Fig.~\ref{fig:rebound}(b)).

        \item The membrane potential then recovers its resting potential on the slow timescale as in the mechanism of (ordinary) excitability (Fig.~\ref{fig:rebound}(c)--(d)).
    \end{itemize}

    If $\Ifast(v;\sigmae)$ intersects the $v$-axis only once, there exists a threshold $\vABE$ such that, for each $v_0<\vABE$, the curves $\Ifast(v;\sigmae)$ and $\Islow(v;v_0)$ intersect at exactly one point.
    Removing an inhibitory current that brings the membrane potential below $\vABE$ thus evokes a rebound spike.
    
    When $\Ifast(v;\sigmae)$ intersects the $v$-axis three times, it also intersects $\Islow(v;v_0)$ three times for every $v_0<\vrest$.
    Thus, the membrane potential recovers to $\vrest$ without generating a spike, regardless of the strength of inhibition.
\end{proof}

In Fig.~\ref{fig:memory}, the number of intersections between $\Ifast(v;\sigmae)$ and the $v$-axis decreases from three to one as $\sigmae$ increases.
By Proposition~\ref{prop:characterization_of_rebound_excitability}, this change gives rise to ABE, provided that the neuron remains excitable at the corresponding noise intensity.
The following proposition shows that this phenomenon occurs for a range of parameter values.

\begin{proposition}
\label{prop:fold_singularity}
    Suppose that $\Ifast(v;0)$ intersects with the $v$-axis three times and that $\bargge(\EEe-\aae)>2\ggl\aae$.
    Then, there exists $\sigmae^{\ast\ast}>0$ such that $\Ifast(v;\sigmae)$ intersects with the $v$-axis
    \begin{itemize}[nosep]
        \item three times if $0\leq\sigmae < \sigmae^{\ast\ast}$,
        \item twice if $\sigmae = \sigmae^{\ast\ast}$,
        \item only once if $\sigmae >\sigmae^{\ast\ast}$.
    \end{itemize}
\end{proposition}

\begin{proof}
    As can be seen in Fig.~\ref{fig:memory}, under the stated parameter condition, $\Ifast(\aae;\sigmae)=-\ggl\aae + \frac{1}{2}\bargge(\EEe-\aae)>0$ for all $\sigmae\geq 0$.
    From the proof of Proposition~\ref{prop:memory}, it follows that, for $v>\aae$, the derivative $\partial_v\Ifast(v;\sigmae)$ either remains negative or is initially positive and then becomes negative as $v$ increases.
    Thus, $\Ifast(v;\sigmae)$ has at most one zero on $(\aae,\infty)$. Since $\Ifast(\aae;\sigmae)>0$ and $\Ifast(\EEe;\sigmae)<0$, it always has one zero in $(\aae,\EEe)$.
    As $\Ifast(v;\sigmae)>0$ for $v<0$, it suffices to focus on the number of zeros of $\Ifast(v;\sigmae)$ on the interval $[0,\aae]$.

    Define $m(\sigmae)\coloneqq \min_{v\in[0,\aae]} \Ifast(v;\sigmae)$.
    We claim that $m(\sigmae)$ increases through zero as $\sigmae$ increases.
    Indeed, since $\Ifast(v;\sigmae)$ is jointly continuous in $(v,\sigmae)$ and nondecreasing in $\sigmae$ for each $v\in[0,\aae]$, $m(\sigmae)$ is continuous and nondecreasing.
    Because $\Ifast(v;0)$ has two distinct zeros on $[0,\aae]$ and is negative between them, we have $m(0)<0$.
    It can be easily shown that $m(\sigmae)>0$ for sufficiently large $\sigmae$.

    Suppose now that $m(\sigma_{\text{e,1}})=m(\sigma_{\text{e,2}})=0$, $\sigma_{\text{e},1} < \sigma_{\text{e},2}$.
    Since $[0,\aae]$ is compact and $\Ifast(0;\sigmae),\Ifast(\aae;\sigmae)>0$, there exist $v_1,v_2\in(0,\aae)$ such that $m(\sigma_{\text{e},i})=\Ifast(v_i;\sigma_{\text{e},i})$, $i=1,2$.
    By the definition of $m(\sigmae)$, we have $\Ifast(v_2;\sigma_{\text{e},1}) \geq  \Ifast(v_1;\sigma_{\text{e},1})=0$.
    Meanwhile, since $\Ifast(v_2;\sigmae)$ is strictly increasing in $\sigmae$, we have $\Ifast(v_2;\sigma_{\text{e},1}) < \Ifast(v_2;\sigma_{\text{e},2})=0$, which is a contradiction.
    Therefore, there exists a unique $\sigmae^{\ast\ast}$ such that $m(\sigmae)<0$ if $0\leq\sigmae<\sigmae^{\ast\ast}$, $m(\sigmae^{\ast\ast})=0$, and $m(\sigmae)>0$ if $\sigmae>\sigmae^{\ast\ast}$.
    
    From the proof of Proposition~\ref{prop:memory}, it follows that $\Ifast(v;\sigmae)$ has at most one local minimum on $(0,\aae)$.
    Together with $\Ifast(0;\sigmae),\Ifast(\aae;\sigmae)>0$, this implies
    that its number of zeros on $[0,\aae]$ is two when $m(\sigmae)<0$,
    one when $m(\sigmae)=0$, and zero when $m(\sigmae)>0$. This completes the proof.
\end{proof}

\begin{figure}
    \centering
    \includegraphics[width=0.97\linewidth]{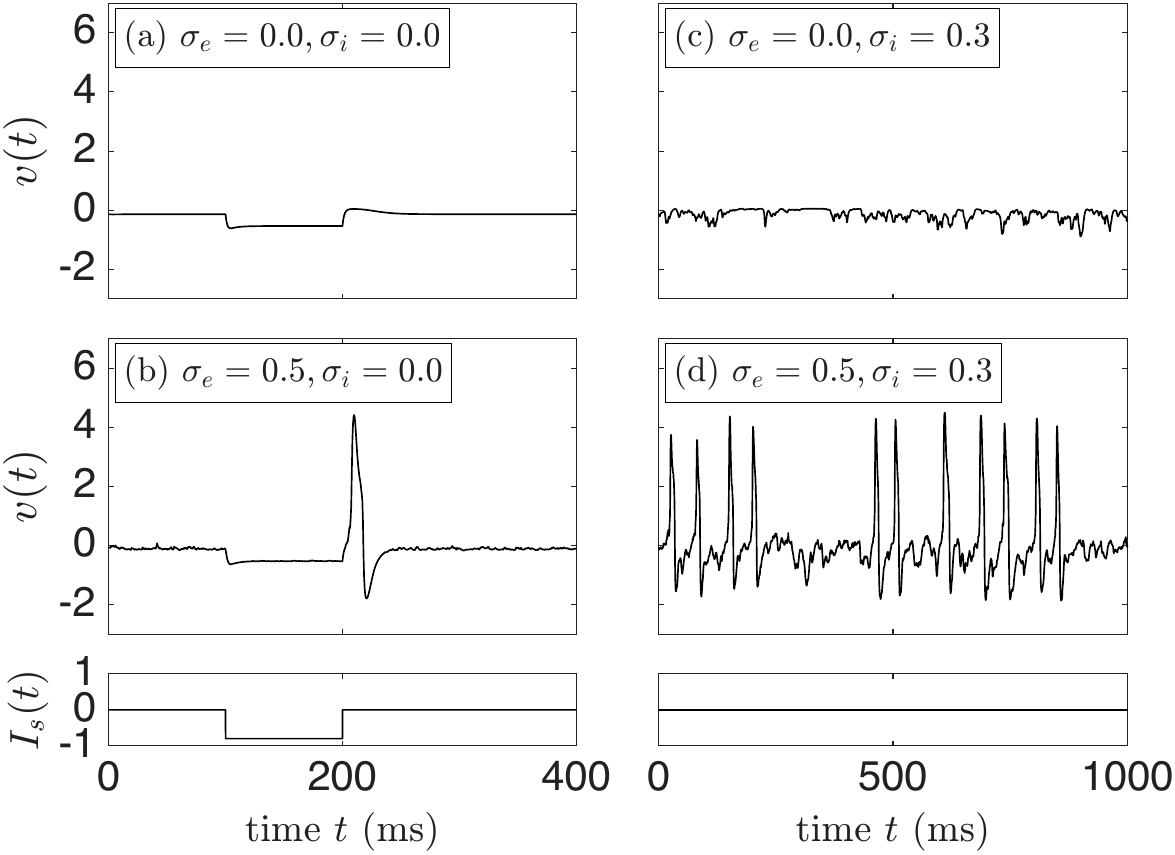}
    \caption{Simulation results for the stochastic memristive neuron \eqref{eqn:stochastic_memristive_model} with parameters listed in Appendix. In panels (a) and (b), an inhibitory external current $\Is=-0.8$ is applied from $t=100$ to $200$.}
    \label{fig:simulation}
\end{figure}

We demonstrate the emergence of ABE through numerical simulations.
Fig.~\ref{fig:simulation}(a)--(b) show responses of the memristive neuron \eqref{eqn:stochastic_memristive_model} to the application and subsequent removal of a prolonged inhibitory current.
The parameters used are listed in Appendix and satisfy the requirements of Proposition~\ref{prop:fold_singularity}.
Without e-noise, the neuron did not exhibit anode break excitation (Fig.~\ref{fig:simulation}(a)).
At an appropriate e-noise intensity, however, it did (Fig.~\ref{fig:simulation}(b)).

\setlength{\textfloatsep}{3mm}
\setlength{\abovecaptionskip}{0.5mm}

\subsection{Quantitative Perspective: Sensitivity}
\label{sec:sensitivity}

So far, we have examined how e-noise qualitatively changes the behavior of the memristive neuron.
In between these qualitative transitions, varying the e-noise intensity modifies $\Ifast$ quantitatively, thereby shifting the resting potential $\vrest$, spike threshold $\vth$, and ABE threshold $\vABE$.
In other words, e-noise also modulates sensitivity.

\begin{figure}
    \centering
    \includegraphics[width=0.93\linewidth]{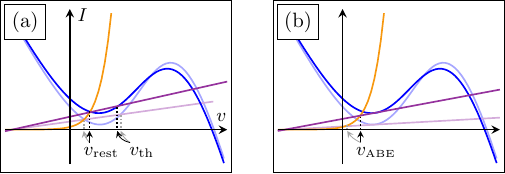}
    \caption{Quantitative effects of e-noise. The curves are defined as in Fig.~\ref{fig:excitability}. The pale curves correspond to a smaller value of $\sigmae$ than the saturated ones.}
    \label{fig:sensitivity}
\end{figure}

Fig.~\ref{fig:sensitivity} illustrates how e-noise modulates sensitivity.
Recall that $\vrest$ can be determined as the intersection of $\Ifast(v;\sigmae)$ and $\Islow(v;v)$, and that $\vth$ corresponds to the middle of the three intersections of $\Ifast(v;\sigmae)$ and $\Islow(v;\vrest)$.
To determine $\vABE$, one can first find the value of $\xxi^\ast$ for which $\Islow(v;\xxi^\ast)$ is tangent to $\Ifast(v;\sigmae)$, as illustrated in Fig.~\ref{fig:sensitivity}(b).
The ABE threshold $\vABE=\xxi^\ast$ can then be read off as the intersection of $\Islow(v;\xxi^\ast)$ and $\Islow(v;v)$.
Now, it can be seen that changing $\sigmae$ reshapes $\Ifast(v;\sigmae)$, thereby shifting these three characteristic potentials.

For the scenario of Fig.~\ref{fig:sensitivity}, one can observe that both $\vth-\vrest$ and $\vrest-\vABE$ decrease as $\sigmae$ increases.
Thus, smaller voltage excursions from rest are needed to reach either $\vth$ or $\vABE$, indicating increased sensitivity to both spike generation and anode break excitation.

\section{Modulation of Stochastic Spiking}
\label{sec:modulation_of_stochastic_spiking}

While e-noise modulates excitability and sensitivity by reshaping $\Ifast$, i-noise can trigger irregular, stochastic spiking. This section investigates how both e- and i-noise modulate this stochastic spiking behavior.

\subsection{Relation to Anode Break Excitability}

We first determine when the (reduced) memristive neuron \eqref{eqn:noisy_reduced_model} exhibits stochastic spiking and when it does not.

\begin{proposition}
\label{prop:characterization_of_random_spiking}
    Consider a memristive neuron governed by \eqref{eqn:noisy_reduced_model}. Suppose that it is excitable in the sense of Definition~\ref{def:excitable}.
    In the presence of i-noise ($\sigmai>0$),
    \begin{itemize}[nosep]
        \item it randomly generates spikes under sufficient timescale separation if $\Ifast(v;\sigmae)$ intersects the $v$-axis only once,
        \item it does not generate spikes under sufficient timescale separation if $\Ifast(v;\sigmae)$ intersects the $v$-axis three times.
    \end{itemize}
\end{proposition}

\begin{figure}
    \centering
    \includegraphics[width=0.93\linewidth]{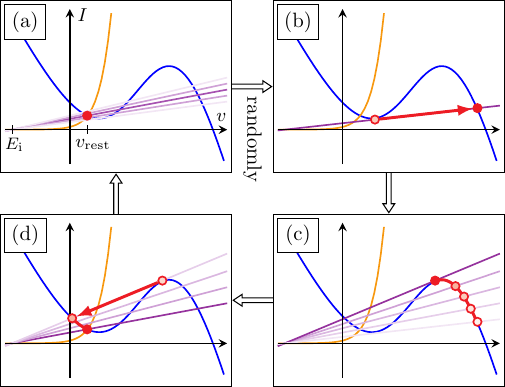}
    \caption{Mechanism of stochastic spiking. The curves are defined as above.}
    \label{fig:random_spiking}
\end{figure}

\begin{proof}
    Suppose that $\Ifast(v;\sigmae)$ intersects the $v$-axis once.
    \begin{itemize}[nosep]
        \item By the i-noise, $\xxi$ fluctuates around the resting potential, causing the slope of $\Islow(v;\xxi)$ to fluctuate, as illustrated in Fig.~\ref{fig:random_spiking}(a).
        Because of the timescale separation, $v$ tracks the intersection of $\Ifast(v;\sigmae)$ and $\Islow(v;\xxi)$.

        \item Eventually, $\xxi$ undergoes a large fluctuation from $\vrest$ that causes $\Ifast(v;\sigmae)$ and $\Islow(v;\xxi)$ to intersect at a single point. (In fact, this occurs when $\xxi<\vABE$.) As in anode break excitation, the fast dynamics then generates a spike (Fig.~\ref{fig:random_spiking}(b)), and the slow dynamics subsequently returns the membrane potential to $\vrest$ (Fig.~\ref{fig:random_spiking}(c)--(d)).
    \end{itemize}

    In contrast, when $\Ifast(v;\sigmae)$ intersects the $v$-axis three times, fluctuations in $\xxi$ cannot trigger a spike. Note that $\Islow(v;\xxi)$ has a positive slope for every value of $\xxi$.
\end{proof}\vspace{-3pt}

Notably, stochastic spiking and ABE arise under the same condition.
This is because both behaviors share a common mechanism:
Sufficient inactivation of the i-current allows the e-current to depolarize the membrane potential.
This finding identifies ABE as a design criterion for regulating stochastic spiking. In particular, ABE is a prerequisite for the stochastic resonance in the considered model, although achieving resonance requires additional tuning of noise intensities.

We demonstrate Proposition~\ref{prop:characterization_of_random_spiking} through numerical simulations.
Using the same parameters as in Section~\ref{sec:rebound}, we introduce i-noise with intensity $\sigmai=0.3$.
Without e-noise, i-noise does not induce spikes (Fig.~\ref{fig:simulation}(c)).
In contrast, i-noise leads to stochastic spiking when e-noise ($\sigmae=0.5$) renders the neuron anode break excitable (Fig.~\ref{fig:simulation}(d)).

\setlength{\abovecaptionskip}{0mm}

\begin{figure*}[t]
    \centering
    \includegraphics[width=0.95\linewidth]{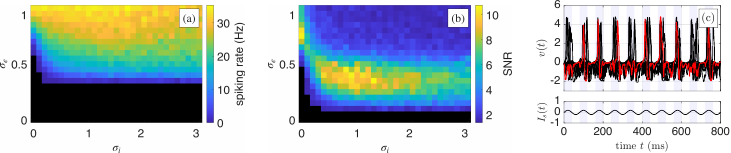}
    \caption{(a) Estimated spiking rate with $\Is(t)=0$ and (b) estimated SNR for the subthreshold input $\Is(t) = 0.2 \sin(2\pi t/80)$ for the memristive neuron \eqref{eqn:stochastic_memristive_model} over a range of $\sigmae$ and $\sigmai$. Black pixels indicate the cases where no spikes were observed. (c) The neuron's responses to the subthreshold input at $\sigmae=0.4$ and $\sigmai=0.6$. Responses from 10 repeated trials are shown, with one trajectory highlighted in red.}
    \label{fig:resonance}\vspace{-3.6mm}
\end{figure*}

\subsection{Sensitivity, Spiking Rate, and Stochastic Resonance}
\label{sec:stochastic_resonance}

We now examine the quantitative modulation of stochastic spiking, beginning with the spiking rate.
Because large fluctuations in $\xxi$ trigger spikes, increasing i-noise is expected to increase the spiking rate. 
Meanwhile, e-noise also affects the spiking rate through a distinct mechanism: by modulating the neuron's sensitivity to anode break excitation.
We numerically estimate the spiking rate of the memristive neuron \eqref{eqn:stochastic_memristive_model}, using the parameters listed in Appendix, over a range of $\sigmae$ and $\sigmai$ values in the absence of external input.
The results are shown in Fig.~\ref{fig:resonance}(a).
Over the parameter range considered, larger values of $\sigmae$ are associated with more frequent spiking.
This is consistent with the observation in Section~\ref{sec:sensitivity} that e-noise increases the sensitivity to anode break excitation.

Stochastic resonance, a noise-induced enhancement of the detectability of a subthreshold periodic signal, is often associated with a match between the frequency of stochastic spiking and the subthreshold signal \cite{gammaitoni1998}.
This suggests that resonance may occur at intermediate noise intensities.
To verify this, we evaluate the signal-to-noise ratio (SNR) for a subthreshold periodic input at $12.5\,\mathrm{Hz}$
over a range of $\sigmae$ and $\sigmai$, following the definition in \cite[p.~228]{gammaitoni1998}.
As shown in Fig.~\ref{fig:resonance}(b), the SNR is maximized at intermediate values of both noise intensities. 
Notably, the spiking rate in the high-SNR region is close to the frequency of the subthreshold input.
Fig.~\ref{fig:resonance}(c) shows the neuron's response at $\sigmae=0.4$ and $\sigmai=0.6$, where the SNR is high.
It can be seen that spikes occur in synchrony with the subthreshold input.

\section{Concluding Remarks}
\label{sec:concluding_remarks}

This paper clarifies how intrinsic noise modulates neuronal behavior, both qualitatively and quantitatively.
We show that e-noise modulates excitability and sensitivity by reshaping the fast current profile, while i-noise triggers stochastic spiking when the neuron exhibits ABE.
Each of these mechanisms is shown to contribute to stochastic resonance in a distinct way.

While the present paper focuses on a specific neuronal model, the established principles may extend beyond it.
For example, the Hodgkin-Huxley model exhibits ABE for the parameter set used in the left panel of Fig.~\ref{fig:HH}, but not for the set used in the right panel.
Future work will extend the analysis to more general conductance-based models and to neuronal networks with synaptic connections.

% \addtolength{\textheight}{-12cm}   % This command serves to balance the column lengths
                                  % on the last page of the document manually. It shortens
                                  % the textheight of the last page by a suitable amount.
                                  % This command does not take effect until the next page
                                  % so it should come on the page before the last. Make
                                  % sure that you do not shorten the textheight too much.

%%%%%%%%%%%%%%%%%%%%%%%%%%%%%%%%%%%%%%%%%%%%%%%%%%%%%%%%%%%%%%%%%%%%%%%%%%%%%%%%

%%%%%%%%%%%%%%%%%%%%%%%%%%%%%%%%%%%%%%%%%%%%%%%%%%%%%%%%%%%%%%%%%%%%%%%%%%%%%%%%

%%%%%%%%%%%%%%%%%%%%%%%%%%%%%%%%%%%%%%%%%%%%%%%%%%%%%%%%%%%%%%%%%%%%%%%%%%%%%%%%
\appendix

The following parameter values are used in the numerical simulations of the memristive neuron \eqref{eqn:stochastic_memristive_model}.
Time is measured in milliseconds, and all other quantities are dimensionless.
\begin{center}
    \setlength{\tabcolsep}{5pt}
    \small
    
    \begin{tabular}{c|c|c|c|c|c|c|c|c|c|c|c}
        \hline
        $C$ & $\ggl$  &  $\bargge$ & $\barggi$ & $\aae$ & $\aai$ & $\bbe$ & $\bbi$ & $\EEe$ &  $\EEi$ & $\epse$ & $\epsi$ \\ \hline
        2 & 1.5 & 1 & 1 & 1.3 & 0.2 & 0.5 & 0.2 & 12 & -5 & 0.03 & 30\\
        \hline
    \end{tabular}

    \vspace{2mm}
\end{center}

\printbibliography

\end{document}